\documentclass[a4paper,11pt]{article}
\usepackage{jinstpub}
\usepackage{inputenc,graphicx}
\usepackage{siunitx}
\usepackage{multirow}
\usepackage{makecell}

\usepackage{lineno}

\newcommand{\dtoa}{\Delta_{\it TOA}}

\title{Time Resolution of a SiGe BiCMOS Monolithic Silicon Pixel Detector without Internal Gain Layer with a Femtosecond Laser}

\author[a]{M. Milanesio,}
\author[a,b]{L. Paolozzi,}
\author[a]{T. Moretti,}
\author[c]{A. Latshaw,}
\author[c]{L. Bonacina,}
\author[a]{R. Cardella,}
\author[a]{T. Kugathasan,}
\author[a,b]{A. Picardi,}
\author[d]{M. Elviretti,}
\author[d]{H. Rücker,}
\author[a]{R. Cardarelli,}
\author[a]{L. Cecconi,}
\author[a]{C. A. Fenoglio,}
\author[a]{D. Ferrere,}
\author[a]{S. Gonzalez-Sevilla,}
\author[a]{L. Iodice,}
\author[a,b]{R. Kotitsa,}
\author[a]{C. Magliocca,}
\author[a,b]{M. Nessi,}
\author[a]{A. Pizarro-Medina,}
\author[a]{J. Sabater Iglesias,}
\author[a]{I. Semendyaev,}
\author[a]{J. Saidi,}
\author[a]{M. Vicente Barreto Pinto,}
\author[a]{S. Zambito,}
\author[a,1]{and G. Iacobucci\note{Corresponding author.}}

\affiliation[a]{Département de Physique Nucléaire et Corpusculaire (DPNC),
University of Geneva, 24 Quai Ernest-Ansermet, CH-1211 Geneva 4, Switzerland}

\affiliation[b]{CERN, CH-1211 Geneva 23, Switzerland}

\affiliation[c]{Département de Physique Appliquée,
University of Geneva, 20 Rue de l'Ecole-de-Médecine, CH-1205 Geneva 4, Switzerland}

\affiliation[d]{IHP — Leibniz-Institut für innovative Mikroelektronik, Im Technologiepark 25, Frankfurt (Oder), Germany}

\emailAdd{giuseppe.iacobucci@unige.ch}

\abstract{The time resolution of the second monolithic silicon pixel prototype produced for the MONOLITH H2020 ERC Advanced project was studied using a femtosecond laser. 
The ASIC contains a matrix of hexagonal pixels with 100 $\mu$m pitch, readout by low-noise and very fast SiGe HBT frontend electronics. 
Silicon wafers with 50 $\mu$m thick epilayer with a resistivity of 350 $\Omega$cm  were used to produce a fully depleted sensor. 
At the highest frontend power density tested of 2.7 W/cm$^2$, the time resolution with the femtosecond laser pulses was found to be 45 ps for signals generated by 1200 electrons, and  3 ps in the case of 11k electrons, which corresponds approximately to 0.4 and 3.5 times the most probable value of the charge generated by a minimum-ionizing particle.
The results were compared with testbeam data taken with the same prototype to evaluate the time jitter produced by the fluctuations of the charge collection.}

\begin{document}
\maketitle
\section{Introduction}\label{sec:into}
Recent research in the framework of the MONOLITH Horizon 2020 ERC Advanced project ~\cite{monolith} demonstrated that SiGe heterojunction bipolar transistor (HBT) electronics can be used to produce low noise, low power, and very fast frontend that could be integrated into a monolithic ASIC to produce a fully efficient detector with excellent time resolution~\cite{Zambito_2023}.
Several ASICs were produced using the 130 nm node SG13G2 SiGe BiCMOS process~\cite{SG13G2} by IHP and characterized at the SPS testbeam facility at CERN with 120 GeV/c pions. 

The  ASIC that included the proof-of-concept PicoAD sensor~\cite{PicoADpatent}, which features a  continuous internal gain layer~\cite{picoad_gain} deep in the silicon sensor bulk, showed time resolutions of 17 ps with minimum-ionizing particles (MIP), with a dependence on the hit position varying from 13 ps at the center of the pixel to 25 ps at the inter-pixel region~\cite{PicoAD_TB}.

More recently, a new prototype was produced with improved electronics. A version without internal gain layer~\cite{Zambito_2023} provided full efficiency and 20 ps time resolution, with a smaller dependence on the position of the hit in the pixel area.
The prototype showed to be radiation tolerant, with a measured time resolution of 40 ps after exposure to a proton fluence of $1\times 10^{16}$ 1 MeV n$_{\text{eq}}$/cm$^{2}$ \cite{Milanesio:radHardSiGe}.

In this paper, we present the results of laboratory measurements performed with a femtosecond laser to assess the timing performance of the SiGe HBT amplifier implemented in the prototype studied in~\cite{Zambito_2023}. The measurement focuses on the characterization of the differential analog outputs, consisting of a fast charge amplifier in SiGe HBT and a two-stage analog driver, that allows for direct measurement of the analog pulses using a fast oscilloscope.

\section{ASIC Description and Characterisation}\label{sec:ASIC}
\begin{figure}[!htb]
    \centering
    \includegraphics[width=.35\textwidth, angle=90]{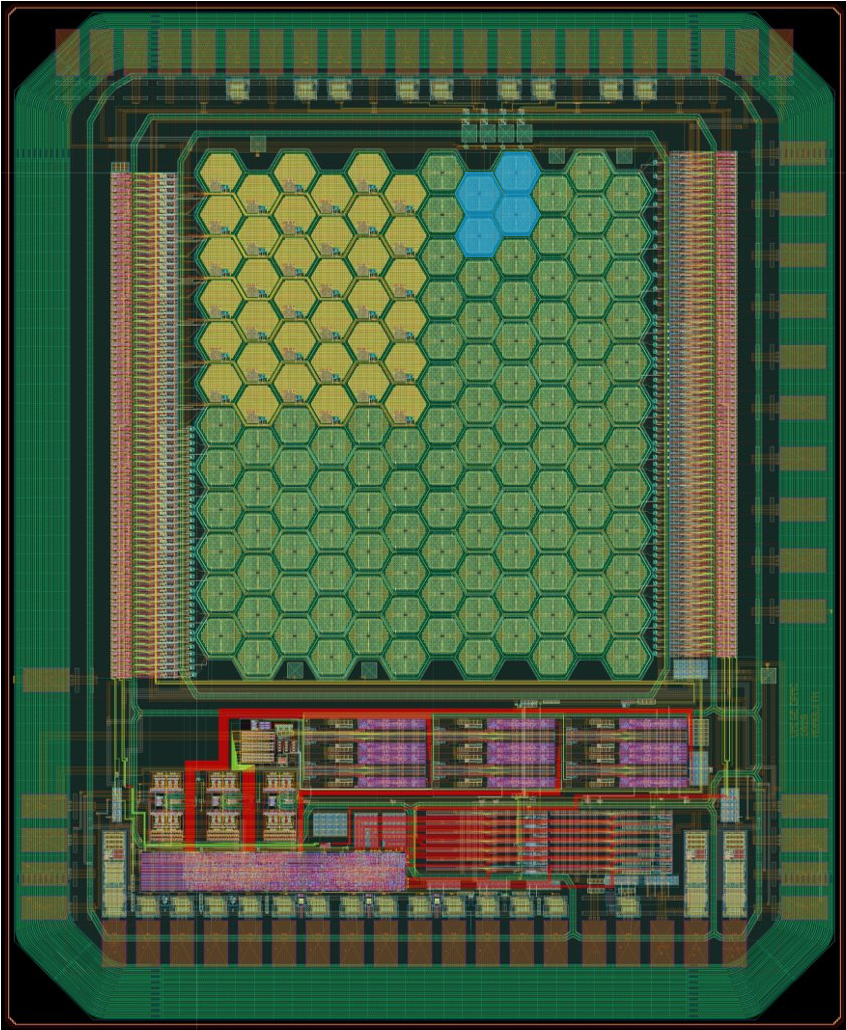}
    \caption{The layout of the monolithic pixel ASIC prototype. The pixel matrix consists of 144 hexagonal pixels of 65 $\mu$m side. Four analog pixels, highlighted in blue, have the preamplification stage directly connected to an analog driver with differential output.
    The yellow color of the bottom-left 6$\times$6 pixels is due to the more dense top-metal lines of this sub-matrix.}
\label{fig:matrix}
\end{figure}

The ASIC  prototype studied here is a monolithic silicon pixel detector in SiGe BiCMOS technology produced using the 130 nm SG13G2 process~\cite{SG13G2} by IHP. It is an evolution of a previous MONOLITH prototype \cite{Iacobucci_2022}, in which the frontend was optimised and its voltage decoupled from the driver to suppress crosstalk between pixels. 
The ASIC matrix, shown in Figure \ref{fig:matrix}, contains 144 hexagonal pixels of 65 $\mu$m side (corresponding approximately to a pixel pitch of 100 $\mu$m), comprising 4 analog pixels with the preamplifier directly connected to an analog driver to be read by an oscilloscope. A more detailed description of the ASIC can be found in \cite{Zambito_2023}.

Two of these ASICs were considered for the measurements with the femtosecond laser. One of them, the reference ASIC, 
provided a very precise time reference, while the other, the detector under test (DUT) ASIC, was used to measure the time resolution as a function of the signal amplitude.
The two ASICs were also characterized in the clean rooms of the University of Geneva using a $^{55}$Fe radioactive source. Table \ref{tab:gain} reports the standard deviation of the electronic-noise distribution of the frontend as measured by the oscilloscope ($\sigma_V$), the preamplifier gain (A$_{q}$), and the equivalent noise charge (ENC) of the reference and the DUT ASICs for different frontend power densities ($P_{\text{density}}$). The ENC was computed as ENC = $\sqrt{\sigma_V^2 -2~\!\sigma_{\it scope}^2}/A_q$, where $\sigma_{\it scope} = (0.16 \pm 0.01)$ mV is the voltage noise of the oscilloscope with the open input connector, taken twice into account because of the two polarities of the differential frontend~\cite{Zambito_2023}.
\begin{table}[!htb]
    \centering
    \renewcommand{\arraystretch}{1.3}
    \begin{tabular}{|c|c|c|c|c|}
    \cline{1-5}
    \cline{1-5}
    \ Chip & $P_{\text{density}}$ [W/cm$^2$] &  $\sigma_V$ [mV] & ~$A_q$ [mV/fC]~ & ENC [electrons] \\
    \cline{1-5}
    ~Reference~ & 2.70 & $ ~1.13 \pm 0.03~ $ & $ ~~74.7 \pm 1.1 $ & $ ~~93\pm 1 $ \\
    \cline{1-5}
    & 2.70 & $ 1.58 \pm 0.04 $ & $ 102.1 \pm 1.2 $ & $ ~~96 \pm 1 $ \\
    DUT & 0.90 & $ 1.06 \pm 0.02 $ & $ ~~67.1 \pm 0.9 $ & $ ~~96 \pm 1 $ \\
    & 0.36 & $ 0.77 \pm 0.02 $ & $ ~~42.7 \pm 0.6 $ & $ 108 \pm 2 $ \\
    \cline{1-5}
    \end{tabular}
    \caption{Voltage noise $\sigma_V$, charge gain $A_q$, and ENC measured with a $^{55}$Fe source for the reference and DUT ASICs,  for different frontend power density. The ASICs were operated at a sensor bias voltage of 200 V.}
\label{tab:gain} 
\end{table}

\section{Femtosecond Laser System}
The \textit{Newport/Spectra-Physics InSight X3} femtosecond laser produced a pulsed light beam at a repetition rate of 80 MHz with less than 200 fs nominal Fourier-Transform-limited pulse duration. The jitter of a single pulse was less than 300 fs and therefore will be considered negligible in the following studies.

The wavelength of the laser was tunable between 680 and 1200 nm. The maximum signal amplitude on the reference chip was obtained for a wavelength of 1042 nm. Therefore, this wavelength was adopted for all the measurements reported here to maximize the charge absorption inside the sensor's active layer.

\section{Experimental Setup and Data Taking}\label{sec:setup}
\begin{figure}[!htb]
\centering
\includegraphics[width=.8\textwidth,trim=0 0 0 0]{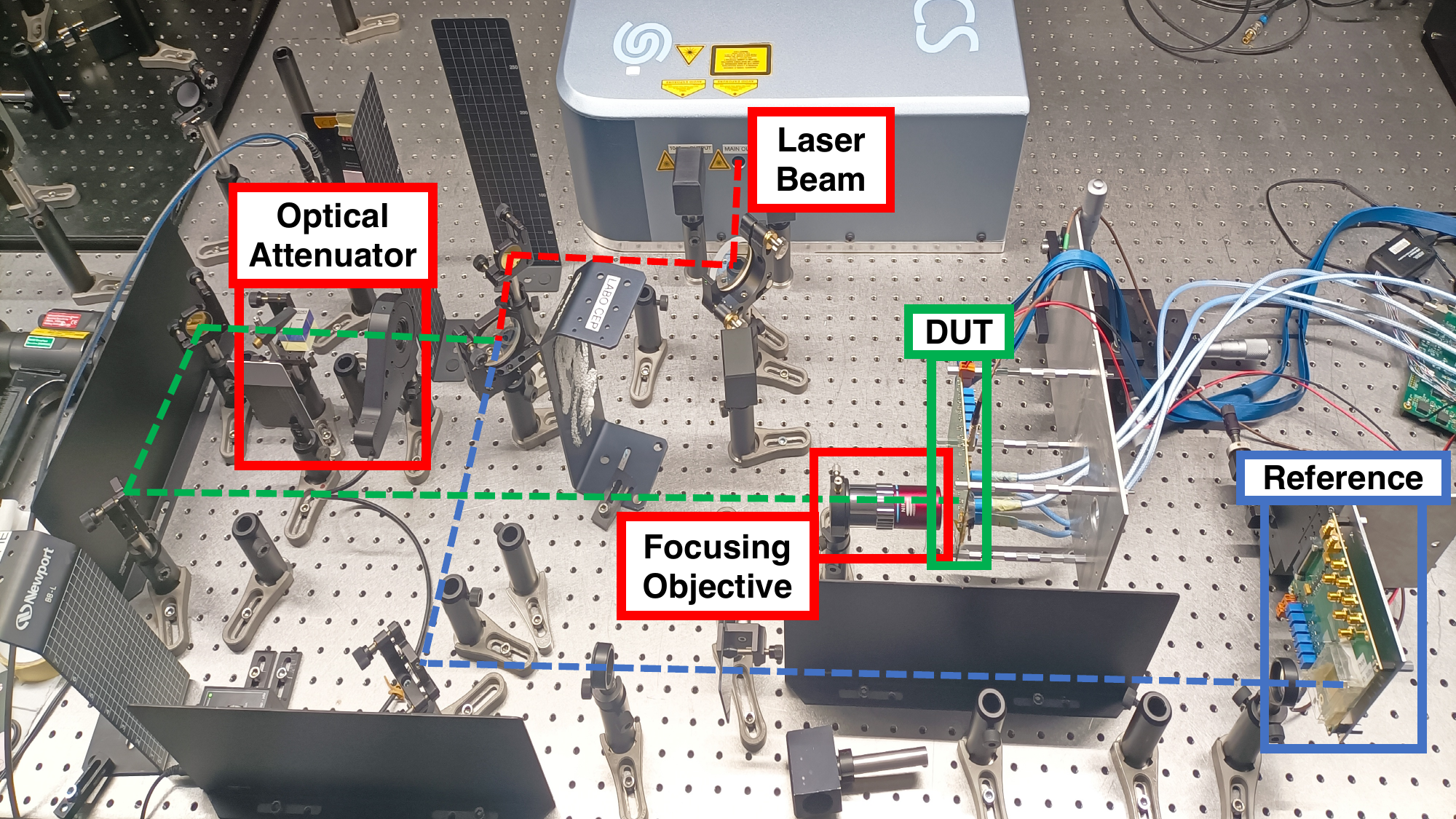}
\caption{The experimental setup. The femtosecond-laser beam was divided into two by a beam splitter, one directed on the DUT ASIC and the second on the reference ASIC. An optical attenuator and a focusing objective were present in front of the DUT to modify the intensity of the laser beam arriving on the ASIC analog pixel studied.}
\label{fig:setup}
\end{figure}

The experimental setup consisted of the femtosecond laser, the reference, and the DUT ASICs, and a 40 GSamples/s  \textit{Lecroy WaveMaster 820Zi-B} oscilloscope connected with high-frequency cables. Figure~\ref{fig:setup} shows the experimental setup (the oscilloscope is not visible).
The beam of the femtosecond laser was divided in two by a splitter, with one beam directed on the reference ASIC and the other beam on the DUT ASIC.

The reference ASIC was always receiving a high-intensity laser beam with a transverse size that covered the entire pixel matrix, centered at the center of the pixel under study. The laser beam saturated the preamplifier and produced large amplitude signals of approximately 230 mV. At $P_{\text{density}}$ = 2.7 W/cm$^2$, this amplitude is reached with 13k electrons, which correspond to four and a half times the most probable value of the charge released by a MIP.
The large signals generated by the reference ASIC in this way provided an excellent time resolution that was used as a precise reference timestamp for the measurement of the time resolution of the DUT as a function of the signal amplitude.

To study the timing performance of the ASIC as a function of the signal amplitude,
the laser beam arriving on the DUT was tightly focused using a focusing objective and centered on one of the four analog pixels.
The intensity of the laser beam arriving on the DUT pixel was varied using an optical attenuator. 

\begin{figure}[!hbt]
    \centering
    \includegraphics[width=0.8\textwidth]{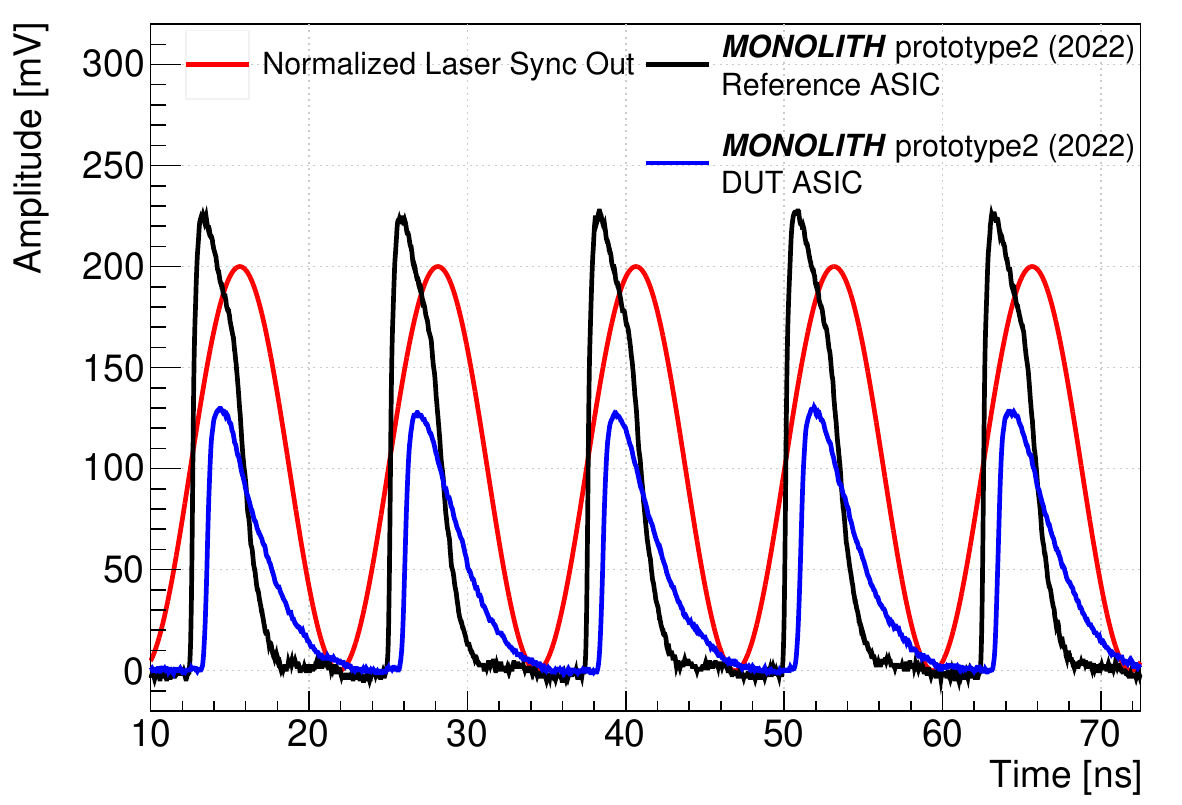}
    \caption{Detail of a waveform acquired using the 40 GSamples/s oscilloscope, that shows the signals produced by five consecutive laser bursts. 
    Each waveform consisted of a train of 40 signals spaced by 12.5 ns.
    The red line shows the laser sync-out signal used to trigger. The signals of the reference ASIC are shown by the black lines, while the signals of the DUT ASIC by the blue lines.  
    This waveform was obtained with the optical attenuator in front of the DUT ASIC regulated to produce signals with an amplitude of 130 mV.
    The much larger laser intensity sent to the reference ASIC produced 230 mV signals that show effects of saturation of the preamplifier.
    }
    \label{fig:waveforms}
\end{figure}

The analog waveforms, each one consisting of a train of 40 consecutive laser pulses, produced by the reference and DUT ASICs were acquired by the oscilloscope together with the laser sync-out signal that was used to trigger.
Figure \ref{fig:waveforms} shows a set of 5 consecutive signals of one of such waveforms. 
The two ASICs are able to operate efficiently and without signal loss with pulses repeated every 12.5 ns.


\section{Results}\label{sec:results}

The time resolution of the DUT with the femtosecond laser was measured starting from the distribution of the difference in time of arrival ($\dtoa$) of the reference and DUT ASICs. The time of arrival was computed as the time at which the signal amplitude reached a constant-fraction threshold of 50$\%$ of the maximum amplitude, calculated offline using a linear interpolation between consecutive signal samplings acquired by the oscilloscope every 25 ps. 
No time-walk correction was applied to the ToA of the signals.

\subsection{Time Resolution of the Reference ASIC with a large laser power}

\begin{figure}[!hbt]
    \centering
    \includegraphics[width=0.8\textwidth]{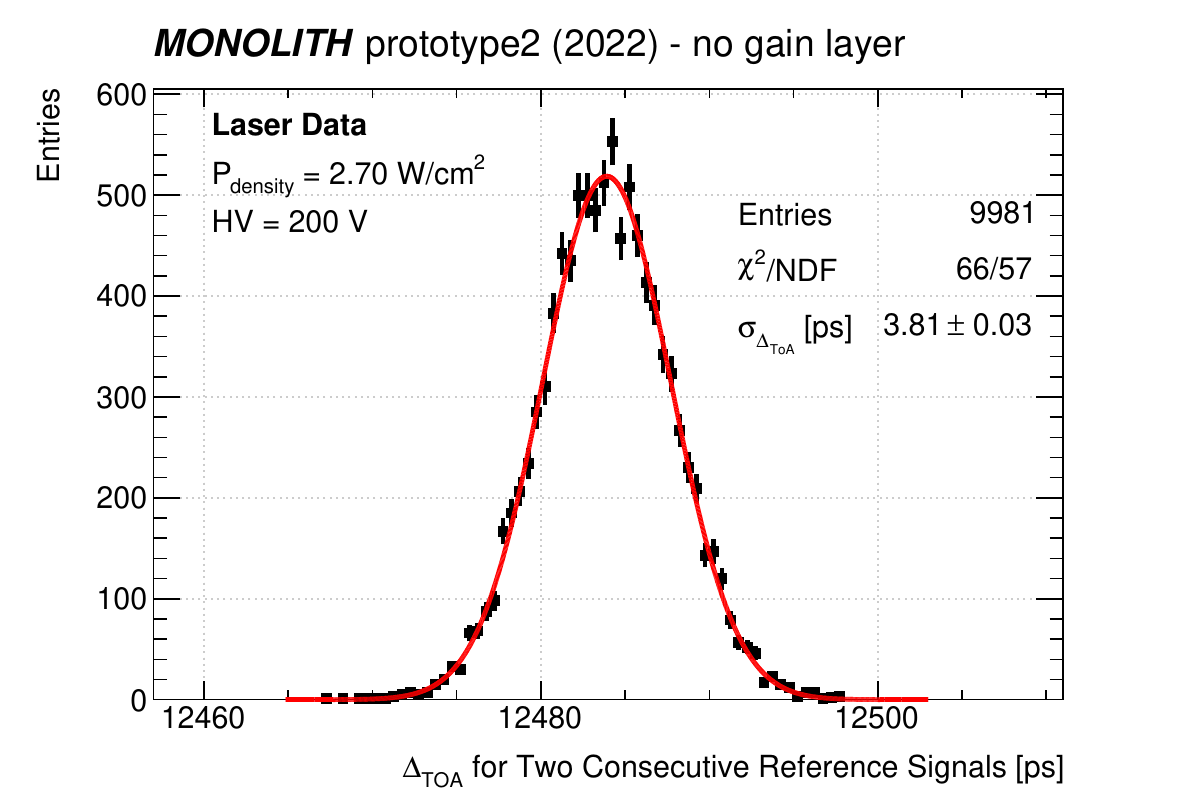}
    \caption{Difference between the ToA of two consecutive signals measured by the reference chip operated at a sensor bias voltage of 200 V and power density of 2.7 W/cm$^2$.
    The red line superposed to the data shows the result of a fit using a Gaussian functional form.}
    \label{fig:deltaToA1signal}
\end{figure}

Figure \ref{fig:deltaToA1signal} shows the distribution of the difference between the ToA of two consecutive signals measured by the reference ASIC. 
The standard deviation of the Gaussian fit of the $\dtoa$ distribution is $\sigma_{\text{$\dtoa$}} = (3.81 \pm 0.03)$ ps. This is the convolution of the time resolution of the reference chip taken two times; hence, the time resolution of the reference chip is $\sigma_t^{\it ref} = (2.69 \pm 0.04)$ ps. 

\subsection{Time Resolution of the DUT vs.  Signal Amplitude Produced by the Laser}
The intensity of the laser on the DUT was varied to acquire signals of different amplitude and therefore different time resolution, while the laser intensity reaching the reference ASIC was kept fixed to generate the signals providing $\sigma_t^{\it{ ref}}$ = 2.7 ps. For each value of signal amplitude in the DUT, approximately 10k waveforms were acquired to be analyzed offline.

\begin{figure}[!htb]
    \centering
    \includegraphics[width=0.8\textwidth,trim=0 0 0 0]{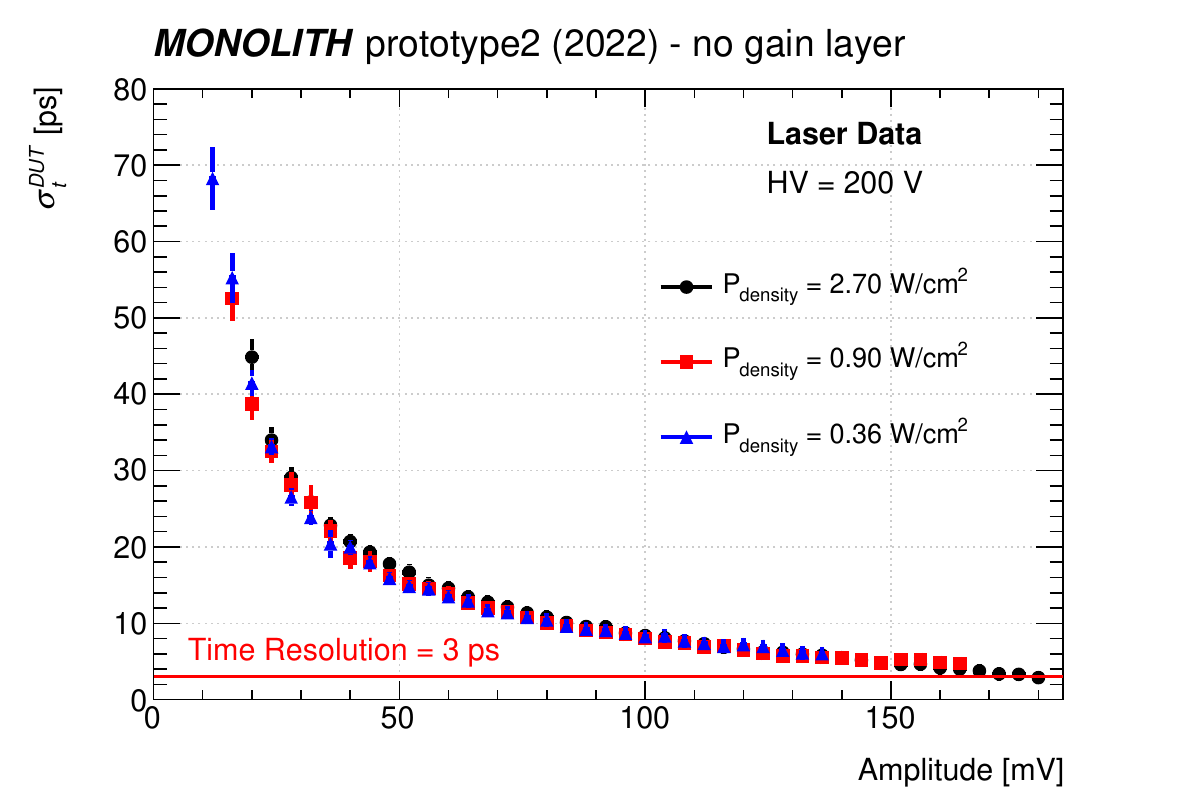}  
    \caption{Time resolution of the DUT as a function of the signal amplitude for the three power density values of Table \ref{tab:gain} at a sensor bias voltage of 200 V.
    The range in amplitude for the three power densities is different since the same set of laser pulse intensities were used for the three measurements.
    }
    \label{fig:sigma_DUT_amp_DUT}
\end{figure}

The time resolution of the DUT was calculated from the distribution of the $\dtoa$ between the reference and the DUT ASICs by subtracting in quadrature the known time resolution  $\sigma_t^{\it ref}$ = 2.7 ps of the reference ASIC, 
$\sigma_t^{\it DUT} = \sqrt{{\sigma_{\text{$\dtoa$}}}^2-2.7^2}$.
The results are reported in figure \ref{fig:sigma_DUT_amp_DUT} as a function of the signal amplitude for the three frontend power densities of Table \ref{tab:gain}.
As expected, the time resolution measured with the femtosecond laser improves as the signal amplitude of the DUT increases. 

At the largest frontend power density studied $P_{\text{density}}$ = 2.7 W/cm$^2$, 
shown by the black dots in figure~\ref{fig:sigma_DUT_amp_DUT}, a time resolution $\sigma_t^{\it DUT}$ = 45 ps is measured for signal  amplitudes of 20 mV,
while $\sigma_t^{\it DUT}$ = 3 ps is reached at 180 mV.
The  gain values reported in Table \ref{tab:gain} can be used to convert these amplitudes into the number of electrons; the result is approximately 1200 and  
11k electrons, respectively. 
Data at larger signal amplitudes, which were used in the case of the reference ASIC to produce the reference timestamp, show 
effects of saturation of the amplifier and, therefore, were not considered in the case of the DUT. 


The time resolutions as a function of the signal amplitude shown in figure \ref{fig:sigma_DUT_amp_DUT} are independent of the operating power of the frontend within uncertainties.
Indeed, in the case of signals generated by a laser beam, the uniform charge distribution inside the sensor volume does not produce additional jitter at large charge values due to charge-collection noise (CCN, also dubbed Landau noise), so the time resolution depends only on the signal amplitude.

In contrast, in the case of a MIP, the non-uniform charge deposition in a sensor would create a CCN that  depends on the total deposited charge; the consequence would be that, for example,
a given amplitude obtained with larger frontend power would imply a smaller electric charge and therefore smaller CCN, and provide better resolution, and the three curves would not overlap.


\subsection{Comparison Between Laser and Testbeam Time Resolutions} 

\begin{figure}[!htb]
     \centering\includegraphics[width=0.8\textwidth]{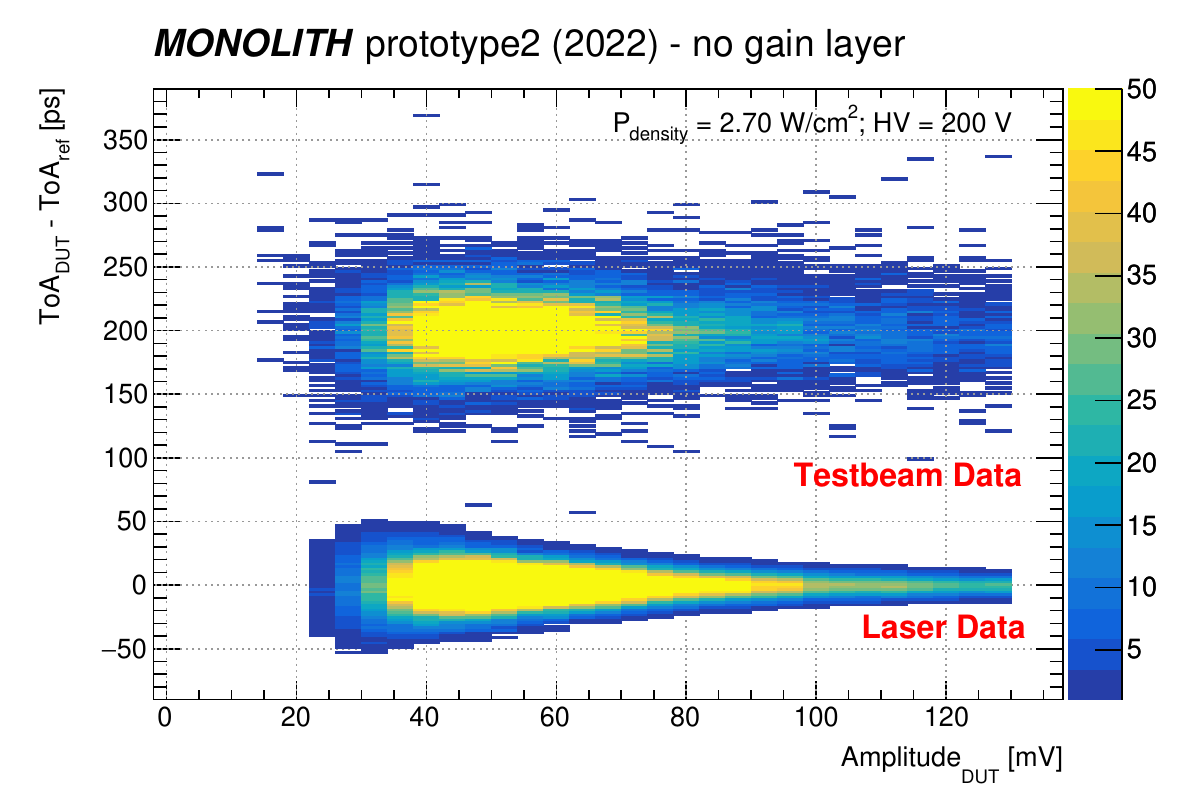}
     \caption{Difference in time of arrival between the reference timestamp and the DUT as a function of the DUT signal amplitude. Both test beam~\cite{Zambito_2023} (top of the plot) and laser measurements (bottom of the plot) were performed at a power density of 2.7 W/cm$^2$ and a sensor bias voltage of 200 V.  For this comparison, the laser data were weighted according to a Landau distribution, and the $\dtoa$ of the testbeam data were shifted by 200 ps to avoid overlap with the laser data.}
     \label{fig:2Dhists}
\end{figure}
\begin{figure}[!htb]
    \centering
    \includegraphics[width=0.8\textwidth]{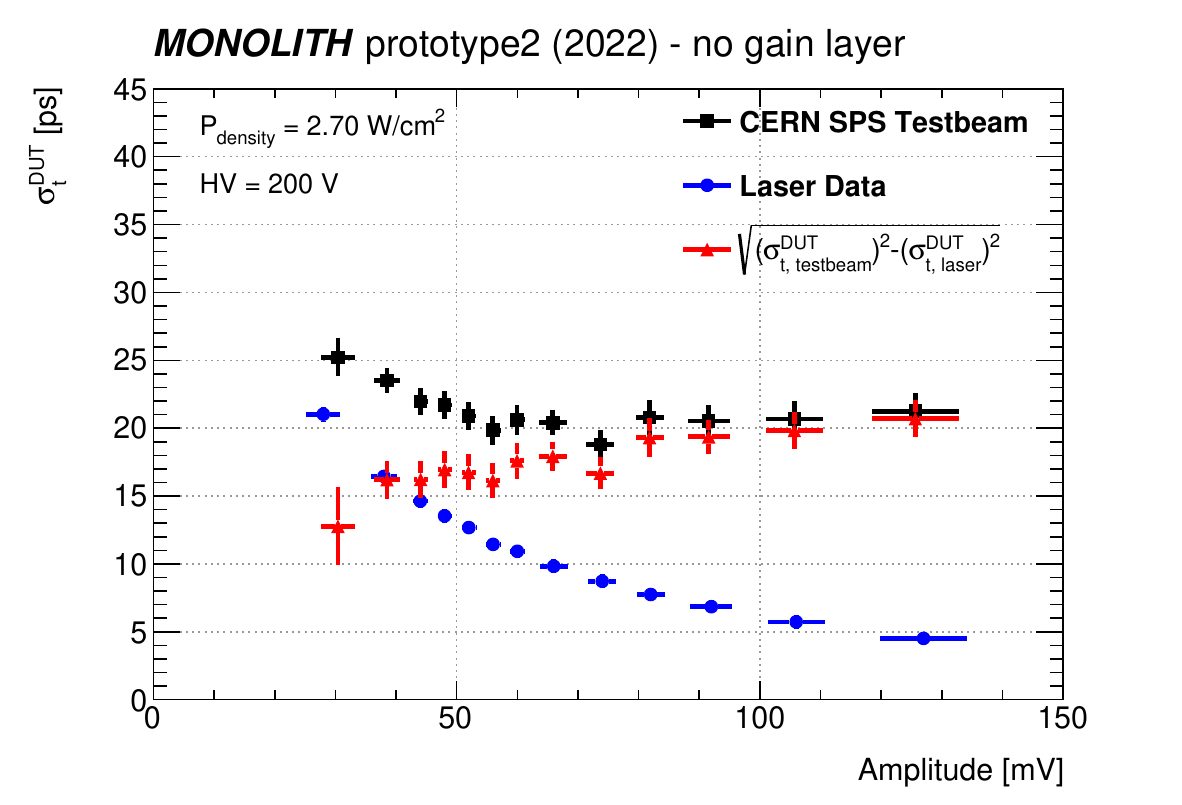}
    \caption{Comparison of the time resolutions as a function of the signal amplitude obtained with the femtosecond laser (blue dots) and with MIPs (black squares) at the CERN SPS testbeam~\cite{Zambito_2023}. The data were taken at a power density of 2.7 W/cm$^2$ and a sensor bias voltage of 200 V. The red triangles show the difference in quadrature between the time resolutions for each bin in amplitude.
    }
    \label{fig:WP1_sigmat_amplitude}
\end{figure}

A comparison of time resolution measurements obtained with MIPs and with a laser could be used to single out the CCN jitter introduced by the fluctuations of the collected charge in the case of a MIP.
This is done in figure \ref{fig:2Dhists}, which reports the difference in time of arrival between the DUT and the reference ASIC produced by the femtosecond laser, shown as a function of the DUT signal amplitude. 
The figure also reports the testbeam data of~\cite{Zambito_2023}, taken with the DUT in the same experimental conditions: sensor bias voltage of 200 V, frontend power density of 2.7 W/cm$^2$, the same region of the pixel (namely accepting only hits within a radius of 30 $\mu m$ from the pixel center\footnote{The laser beam profile was measured to have a radius of approximately 30 $\mu$m using an array of photodiodes. The testbeam data were therefore restricted to tracks reconstructed by the testbeam-telescope that were crossing the sensor within 30 $\mu$m from the pixel center.}), and the same settings of the oscilloscope. 
In the case of the testbeam data, the reference timestamp was provided by two MCP detectors, each with a time resolution of 5 ps, and the data were corrected for signal time walk~\cite{Zambito_2023}. 
The width of the $\dtoa$ distribution of the laser data gets smaller for higher amplitudes, as already found in figure \ref{fig:sigma_DUT_amp_DUT}. In contrast, the width of the distribution of the testbeam data remains rather large at large amplitudes. This difference could be attributed to the presence in the MIP data of the noise coming from the charge-collection.
To quantify this difference, figure \ref{fig:WP1_sigmat_amplitude} shows the time resolution as a function of the signal amplitude obtained with the testbeam data described in~\cite{Zambito_2023} (black squares) and with the laser (blue dots).
In the case of the testbeam data, the signal amplitudes were limited to 140 mV by the oscilloscope readout.
In both cases, for each bin of signal amplitude, the time resolution of the DUT was obtained by subtracting in quadrature the timestamp of the reference detector from the $\sigma_{\text{$\dtoa$}}$.

\subsection{Extraction of the Charge-Collection Noise Contribution} 
The red triangles in figure \ref{fig:WP1_sigmat_amplitude} show the difference in quadrature between the time resolutions measured at the testbeam and with the laser and provide a measurement of the time jitter produced in the testbeam data by the CCN as a function of the signal amplitude.
This result shows that the electronics jitter dominates the time resolution at low signal amplitudes, while the CCN dominates the time resolution at high signal amplitudes. It also shows that the CCN contribution is not constant, but it increases when the signal amplitude increases.

\begin{figure}[!htb]
     \centering\includegraphics[width=0.8\textwidth]{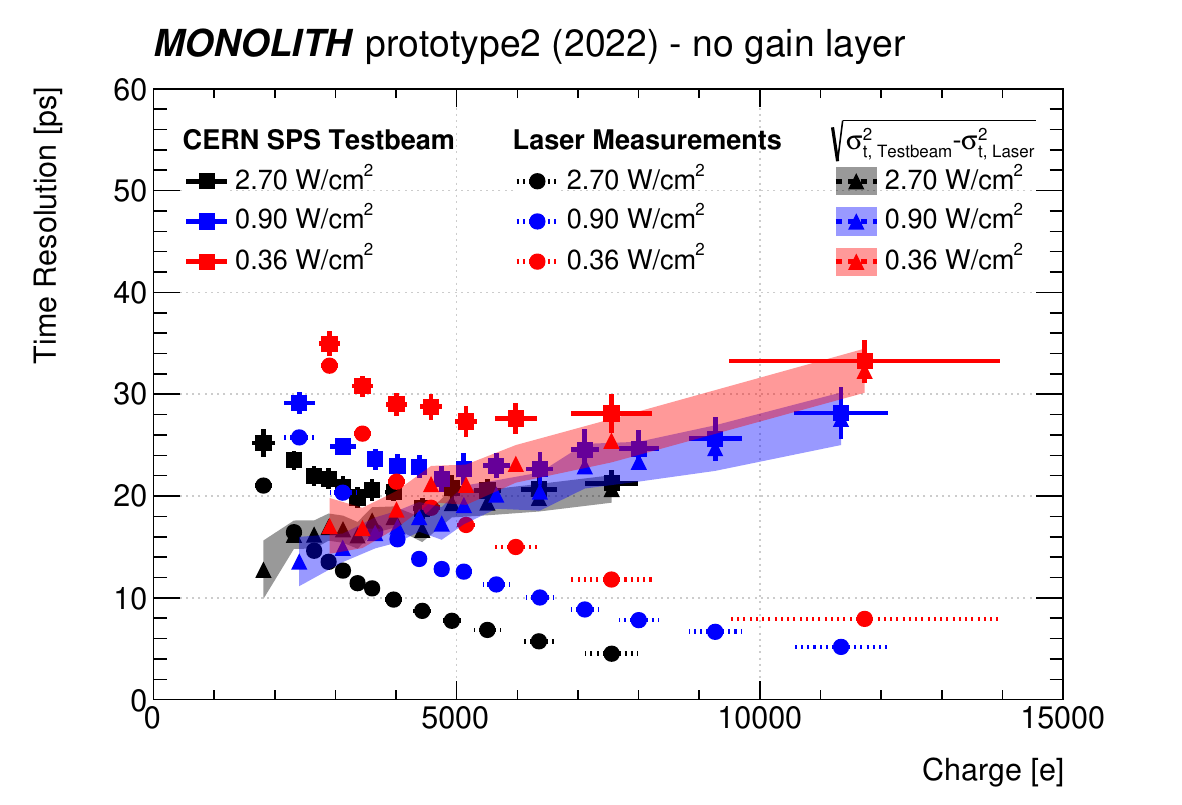}
     \caption{Time resolutions as a function of the charge injected in the frontend electronics, obtained with MIPs at the CERN SPS testbeam~\cite{Zambito_2023} and with the femtosecond laser. The data were taken at a sensor bias voltage of 200 V and at the three power densities listed in Table \ref{tab:gain}.
     The bands show the difference in quadrature between the time resolutions for each bin of charge, the width corresponding to one standard deviation. }
     \label{fig:total_comparison}
 \end{figure}

To gather more information on the time jitter produced by the CCN, it is instructive to compare the laser and the testbeam data for the three power densities listed in Table~\ref{tab:gain} as a function of the electric charge, instead of the amplitude. This is done in
Figure \ref{fig:total_comparison}, which presents the laser and testbeam resolutions as well as their difference in quadrature. The injected charge in the frontend electronics was obtained by dividing the amplitude of the signal by the gain for each power density reported in Table~\ref{tab:gain}.
Only signals with amplitudes less or equal to 140 mV were considered for this study, since the testbeam signals with amplitude beyond that value were saturated by the oscilloscope setting and therefore could not be time-walk corrected with sufficient accuracy. 
For this reason the data displayed as a function of the injected charge have different ranges, and the data taken at larger power density of the frontend result to be limited to smaller charge values.

Both laser and testbeam data in figure~\ref{fig:total_comparison} show better time resolutions when the power density of the frontend electronics is higher.
On the other hand, the two sets of data behave very differently as a function of the charge: 
while the time resolutions of the laser data improve strongly at large charge values, the testbeam data show time resolutions that get worse at large charge values. This observation can be explained by the presence of the time jitter due to the CCN, which was calculated as the difference in quadrature and is represented by the colored bands.
For the three power density values, the CCN time jitter shows a clear increase as a function of the charge producing the signals and dominates the time resolution at the largest charge values measured.
The red band showing the CCN contribution at 0.36 W/cm$^2$ seems to rise more quickly than the other two bands, although this observation is marginal due to the fact that the band limits represent one standard deviation uncertainty of the measurement.

\section{Conclusions}\label{sec:conclusions}
The time resolution of the monolithic silicon pixel prototype without an internal gain layer produced for the H2020  ERC Advanced MONOLITH project was studied using pulses from a femtosecond laser. The prototype was operated at sensor bias voltage of 200 V and frontend power density between 0.36 and 2.7 W/cm$^2$. At the highest power density, the time resolution was found to vary between 45 ps and 3 ps for laser pulses producing amplitudes between 20 and 180 mV, corresponding approximately to 0.4 and 3.5 times the most probable value of the charge deposited by a minimum-ionizing particle, respectively.  Comparison with the time resolution obtained at the same working point with data taken at a testbeam with minimum-ionizing particles shows that the time jitter due to the charge-collection noise increases with the signal amplitude and dominates the time resolution at large signal amplitudes.

\acknowledgments
This research is supported by the Horizon 2020 MONOLITH  ERC Advanced Grant ID: 884447. The authors wish to thank Coralie Husi, Javier Mesa, Gabriel Pelleriti, and the technical staff of the University of Geneva and IHP Microelectronics.
The authors acknowledge the support of EUROPRACTICE in providing design tools and MPW fabrication services.

\newpage
\bibliographystyle{unsrt}
\bibliography{bib.bib}

\end{document}